\documentclass[sigconf]{acmart}

\copyrightyear{2025}
\acmYear{2025}
\setcopyright{acmlicensed}
\acmConference[WWW Companion '25] {Companion of the 16th ACM/SPEC International Conference on Performance Engineering}{April 28-May 2, 2025}{Sydney, NSW, Australia.}
\acmBooktitle{Companion of the 16th ACM/SPEC International Conference on Performance Engineering (WWW Companion '25), April 28-May 2, 2025, Sydney, NSW, Australia}
\acmISBN{979-8-4007-1331-6/25/04}
\acmDOI{10.1145/3701716.3715212}

\usepackage{graphicx}
\usepackage{natbib}
\usepackage{caption} 
\usepackage{todonotes}
\usepackage{booktabs}
\usepackage{algorithm}
\usepackage{algorithmic}
\usepackage{newfloat}
\usepackage{listings}
\usepackage{adjustbox}
\usepackage[title]{appendix}

\DeclareMathOperator{\coocur}{co\_occur}
\DeclareMathOperator{\len}{len}
\DeclareMathOperator{\edist}{edit\_dist}
\DeclareMathOperator{\escore}{edit\_score}

\begin{document}

\title{ClarAVy: A Tool for Scalable and Accurate Malware Family Labeling}

\author{Robert J. Joyce}
\orcid{0009-0003-7168-1237}
\affiliation{%
  \institution{Booz Allen Hamilton}
  \city{McLean}
    \state{VA}
  \country{USA} 
}
\email{joyce8@umbc.edu}

\author{Derek Everett}
\orcid{0000-0003-3593-5255}
\affiliation{%
  \institution{Booz Allen Hamilton}
  \city{McLean}
  \state{VA}
  \country{USA} 
}
\email{Everett_Derek@bah.com}

\author{Maya Fuchs}
\orcid{0000-0002-0771-2647}
\affiliation{%
  \institution{Booz Allen Hamilton}
  \city{McLean}
  \state{VA}
  \country{USA} 
}
\email{Fuchs_Maya@bah.com}

\author{Edward Raff}
\orcid{0000-0002-9900-1972}
\affiliation{%
  \institution{Booz Allen Hamilton}
  \city{McLean}
  \state{VA}
  \country{USA} 
}
\email{Raff\_Edward@bah.com}

\author{James Holt}
\orcid{0000-0002-6368-8696}
\affiliation{%
  \institution{Laboratory for Physical Sciences}
  \city{College Park}
  \state{MD}
  \country{USA} 
}
\email{holt@lps.umd.edu}

\renewcommand{\shortauthors}{Robert J. Joyce, Derek Everett, Maya Fuchs, Edward Raff, \& James Holt}

\begin{abstract}
Determining the family to which a malicious file belongs is an essential component of cyberattack investigation, attribution, and remediation. Performing this task manually is time consuming and requires expert knowledge. Automated tools using that label malware using antivirus detections lack accuracy and/or scalability, making them insufficient for real-world applications. Three pervasive shortcomings in these tools are responsible: (1) incorrect parsing of antivirus detections, (2) errors during family alias resolution, and (3) an inappropriate antivirus aggregation strategy. To address each of these, we created our own malware family labeling tool called ClarAVy. ClarAVy utilizes a Variational Bayesian approach to aggregate detections from a collection of antivirus products into accurate family labels. Our tool scales to enormous malware datasets, and we evaluated it by labeling $\approx$40 million malicious files. ClarAVy has 8 and 12 percentage points higher accuracy than the prior leading tool in labeling the MOTIF and MalPedia datasets, respectively.
\end{abstract}



\maketitle

\section{Introduction}
\label{sec:introduction}

A malware family is a collection of malicious files originating from the same source code \cite{MOTIF}. Over time, malware authors update their malware source code to add new functionality and to evade detection, resulting in new instances of the family that are behaviorally similar to previous versions \cite{wadkar2020}. Other techniques, such as polymorphism and packing, are also responsible for creating derivative files belonging to the same family \cite{osorio2015, ucci}. 
Identifying the family to which a malicious file belongs is challenging, but provides valuable information to analysts and incident responders \cite{bayer}. 
Unfortunately, doing this manually is time-consuming -- an analyst may take hours or even days to fully analyze a malware specimen \cite{mohaisen2013,247696}.

\begin{figure}[t!]
    \centering

\includegraphics[width=0.85\columnwidth,keepaspectratio]{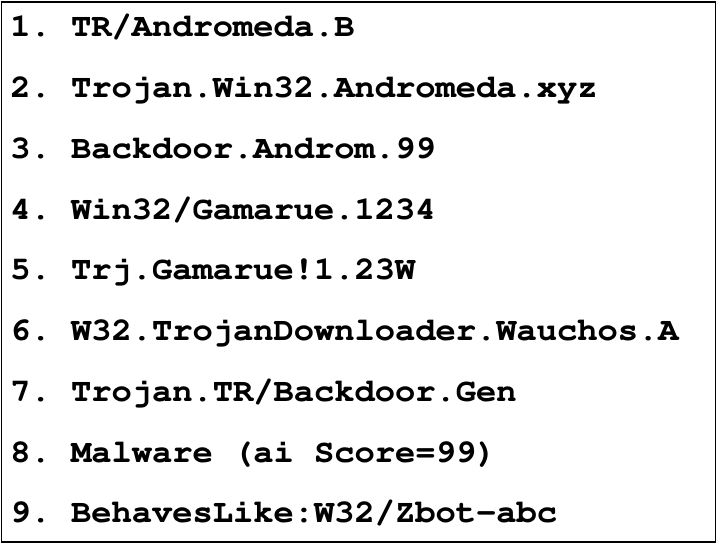}
    \caption{Fictitious AV scan report with detections from 9 AV products. AV products 1-6 correctly detect that this file belongs to the Andromeda family, which has the aliases Androm, Gamarue, and Wauchos. AV 7 predicts that the malware is a trojan and a backdoor. AV 8 uses an ML-based approach to detect the file as malicious. The heuristic used by AV 9 results in an incorrect detection as the Zbot family.}
    \label{fig:scan-results}
\end{figure}

Automating malware family classification is an active area of research \cite{bayer, nataraj, mohaisen2015, huang, ucci}. Family labels for large malware datasets are typically derived from antivirus (AV) detections \cite{joyce2023maldict, bodmas, vxheaven}. In particular, the leading approaches label malware by aggregating the detections of multiple AV products (which we refer to as AV scan reports, see Figure \ref{fig:scan-results}), rather than relying on any individual AV product \cite{botacin, mohaisen2014}.
We developed a malware labeling tool named ClarAVy (pronounced "clarify") that aggregates AV scan reports into the following information:

\begin{itemize}
    \item The hash of the scanned file.
    \item The number of AV products that detected the file as malicious and the total number of AVs that scanned the file.
    \item The most likely malware family and a confidence score for the prediction.
    \item A list of tags applied to the file.
\end{itemize}

This paper builds upon our earlier work on ClarAVy \cite{joyce2023maldict}. The primary contribution of this work is to add functionality for intelligent and scalable malware family labeling to our tool. We have also made incremental contributions to ClarAVy's parsing, alias resolution, and tagging modules.

\section{Related Work} \label{sec:related_work}

We now review five tools that use AV scan reports to label malware. \textit{AVClass} is the seminal work in this area \cite{avclass}. The tool has a module for generating a token taxonomy and identifying aliases, although the module is not enabled by default. AVClass's successor, \textit{AVClass2}, adds functionality to tag malware according to file properties, behaviors, and other malicious attributes \cite{avclass2}. AVClass and AVClass2 have since been combined into the same codebase and are collectively referred to as "AVClass" in this work \cite{avclassgithub}. \textit{EUPHONY} is an AV tagging tool that was developed for labeling Android malware in particular. It uses a graph-based community detection approach for labeling clusters of similar files \cite{euphony}. \textit{AVClass++} is a fork of the original AVClass tool (prior to the merge with AVClass2) that can perform label propagation \cite{avclassplusplus}. \textit{Sumav} automatically generates a token taxonomy by forming a graph of related tokens, from which it identifies families \cite{sumav}. \textit{TagClass} uses an incremental parsing strategy that accurately identifies family names appearing after behavior- and file-related tokens \cite{tagclass}. In subsequent work, the authors of TagClass were the first to propose a family labeling approach based on the Dawid-Skene algorithm \cite{dawid-skene, tagclass_DS}.
The code for TagClass' AV parsers is available online, but the authors have not yet published an implementation of their labeling methodology based on the Dawid-Skene algorithm.

Accurate malware labeling is a critical issue in both research ~\cite{TirthCAMLIS} and industry ~\cite{10.1145/3097983.3098196} settings. We found that each of the surveyed tools makes crucial oversights in one or more of the following areas, leading to labeling errors.

    (1) \textbf{AV Detection Parsing}. Special care must be taken to avoid mistakes when parsing AV detections. Improper parsing can result in family names being ignored or non-family tokens being treated as family names.
    
    (2)  \textbf{Alias Resolution}. Each AV product uses their own preferred naming conventions. Tokens with different spellings but the same meaning are called \emph{aliases} and must be identified. Existing tools make frequent aliasing errors or ignore the alias problem entirely. 
    
    (3) \textbf{Aggregation Strategy.} Individual AV products may have better or worse coverage of particular malware families, and correlations between AV products must be taken into account. Existing tools use aggretation strategies that are na\"ive or that do not scale to large dataset sizes.

\section{ClarAVy AV Parsing}
\label{sec:claravy-parsing}

In this section, as well as Sections \ref{sec:alias} and \ref{sec:family-labeling}, we discuss ClarAVy's improvements in AV detection parsing, alias resolution, and aggregation strategy. We begin with ClarAVy's AV parsing implementation. ClarAVy parses AV detections by tokenizing them and then assigning each token to one of the lexical categories in Table \ref{tab:taxonomy}. Since our prior work, ClarAVy's parsers have been updated to support threat group tokens, and we have added special handling for family tokens \cite{joyce2023maldict}.

ClarAVy tokenizes AV detections by separating them on non-alphanumeric characters. For example, ClarAVy would separate the AV detection Exploit:Win32/MS08067.xyz into the tokens "Exploit", "Win32", "MS08067", and "xyz". ClarAVy also identifies the structure of each AV detection. As an example, the structure of Exploit:Win32/MS08067.xyz is TOK:TOK/TOK.TOK, where "TOK" indicates the location of a token in the AV detection.

\begin{table}[!h]
\centering
\caption{ClarAVy's token taxonomy. Every token is assigned to one of these 10 lexical categories. For example, in the AV detection Trojan.Win32.Andromeda.xyz,   "Trojan" is a BEH token, "Win32" is a FILE token, "Andromeda" is a FAM token, and "xyz" is a SUF token. }
\label{tab:taxonomy}

\begin{tabular}{@{}l@{\hskip 12pt}l@{}}
\toprule
FAM & The malware family that the file belongs to \\
\midrule
GRP & An APT group, campaign, or operation \\
\midrule    
BEH & The malware category or a malicious behavior \\
\midrule
FILE & The OS, file format, or programming language \\
\midrule
VULN & A vulnerability exploited by the malware\\
\midrule
PACK & The packer used to pack the file \\
\midrule
HEUR & Indicates the AV detection is a heuristic\\
\midrule
SUF & A suffix token at the end of the AV detection \\
 \midrule
PRE & Ambiguous, but not a FAM or SUF token \\
 \midrule
UNK & Used in the rare case of  truly ambiguous tokens \\
\bottomrule
\end{tabular}
\end{table}

Detections from the same AV product and with the same structure can generally be parsed in a predictable manner. For example, for all detections with the TOK:TOK/TOK.TOK structure that were output by a particular AV product, we found that the first token always indicated a malicious behavior, the second token always indicated a file property, and the fourth token was always a "suffix". The third token varied; it was either the name of a malware family or a vulnerability that the file exploited.

\begin{figure}[!h]
    \centering
\includegraphics[width=0.975\columnwidth,keepaspectratio]{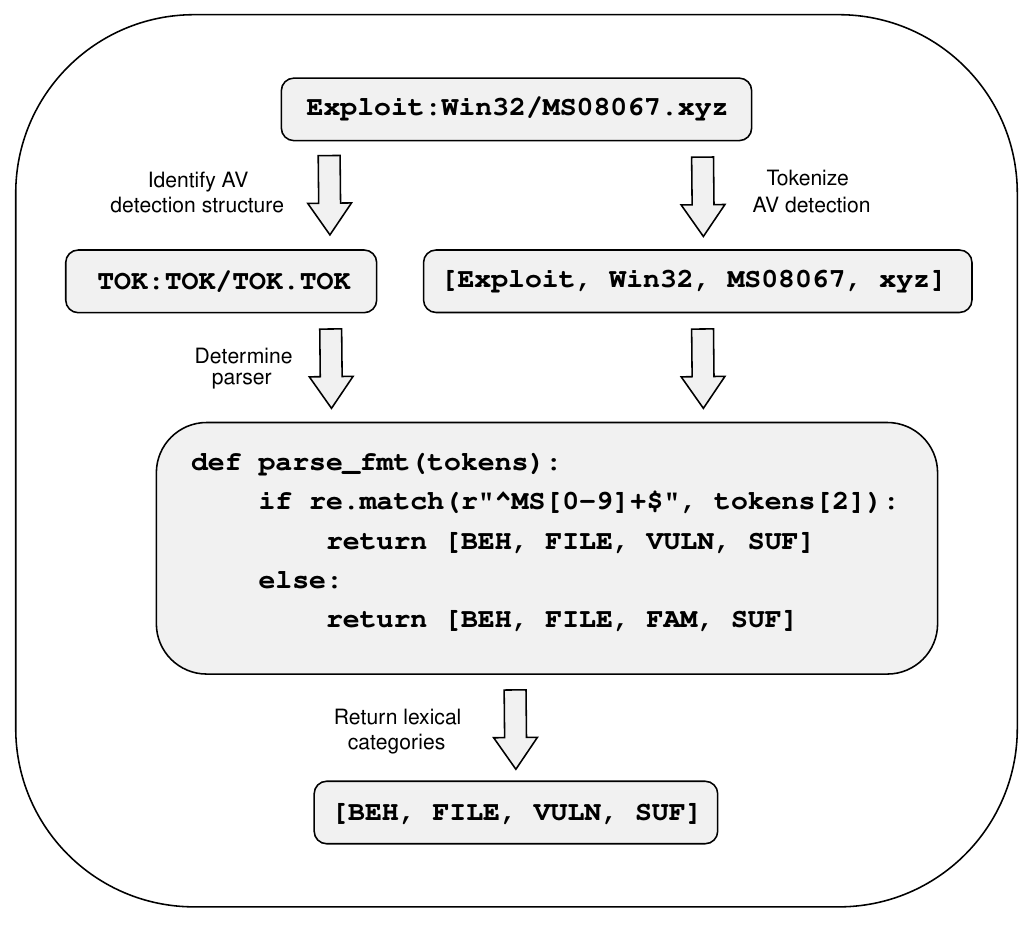}

    \caption[Parsing of an Antivirus Detection]{Parsing  the detection Exploit:Win32/MS08067.xyz. ClarAVy tokenizes it and selects a parsing function based on the structure TOK:TOK/TOK.TOK. A regular expression in the parsing function finds that the third token is the MSO8-067 vulnerability, rather than a family name.}
    \label{fig:parsing}
\end{figure}

After tokenizing an AV detection, ClarAVy identifies an appropriate parsing function based on the AV detection's structure. These parsing functions work by applying regular expressions and/or Boolean logic to the tokens. The parsing function then returns a lexical category assignment for each token in the AV detection. Figure \ref{fig:parsing} shows an example of an AV detection being parsed.

At present, ClarAVy includes 934  \textit{manually-crafted} parsing rules for the most common AV detection structures used by 103 notable AV products. We have added support for an additional 14 AV products since our prior work \cite{joyce2023maldict}. The parsing functions range from simple -- as in the example shown in Figure \ref{fig:parsing} -- to complex, depending on how consistent the detections of an AV product are.  
To create these parsing functions, we collected 39,747,485 VirusTotal reports for files in chunks 0-465 of the VirusShare corpus \cite{virusshare}. Each parsing function was developed and validated using at least 10,000 AV detections with the corresponding structure. Our parsing functions have coverage for over 99.5\% of the 1.1 billion AV detections in this set of VirusTotal reports \cite{joyce2023maldict}.

\subsection{Handling Parsing Ambiguity}

In some cases, there are no means for identifying a token's lexical category using pattern-matching. The most common instance of this is when file-related, behavior-related, and/or other generic tokens may be found in the same location of an AV detection, with no means of distinguishing them. In these cases, ClarAVy assigns the PRE lexical category (see Table \ref{tab:taxonomy}) to denote that there is some ambiguity, but the token is not a family name or a suffix token. More rarely, there are edge cases where tokens are truly ambiguous. ClarAVy uses the UNK lexical category for these tokens.

After the entire set of AV scan reports have been parsed for the first time, ClarAVy inspects each token and attempts to assign it to a permanent lexical category. This is done by enumerating which lexical categories a token was assigned during parsing. If a token was almost always assigned a single lexical category (excluding PRE and UNK), it becomes permanently associated with that category. 
For example, if the parsing functions assigned Backdoor token BEH 1,000 times, PRE 200 times, and FAM 5 times, it would be permanently associated with the BEH lexical category. If an AV detection containing the Backdoor token is parsed later, it would immediately be known to be a behavior-related token.

\subsection{Special Token Handling}
\label{sec:special-cases}

Special handling is required for certain types of tokens. For example, tokens that are the names of threat groups are typically co-located with family tokens in AV detections. Although they can sometimes be identified using pattern matching, GRP and FAM tokens often cannot be distinguished. We manually assembled a list of over 250 known APT group, operation, and campaign names to identify GRP tokens more reliably. Another case of special handling is for the names of vulnerabilities, which are often sequences of multiple tokens (e.g. "CVE\_2017\_0144" is tokenized into three tokens). ClarAVy uses regular expressions and other pattern matching logic to identify these spread-out vulnerability names. 

Malware family names are treated more strictly than other types of tokens to ensure parsing accuracy. If a FAM token is used by fewer than three unrelated AV products, and it is not found to be the alias of another family, it is downgraded to an UNK token.
In addition, it is common for AV products to use placeholder families when they are unable to determine a more specific family name. For instance, Razy, WisdomEyes and Artemis are placeholder family names used by Bitdefender, Baidu, and McAfee, respectively \cite{hahn}. We manually curated a list of common placeholder families, and ClarAVy ignores them.

\section{ClarAVy Alias Resolution}
  \label{sec:alias}

Once the token taxonomy has been built, ClarAVy attempts to handle aliases; tokens that have different spellings but identical meanings. We identify three different classes of token aliases, which we call \textbf{trivial aliases}, \textbf{sibling aliases}, and \textbf{parent-child aliases}. This section describes ClarAVy's approach for identifying and resolving each of these types of aliases. Trivial alias resolution (see Section \ref{sec:trivial-alias}) and parent-child alias resolution (see Section \ref{sec:parent-child}) for non-family tokens originates from our prior work on ClarAVy \cite{joyce2023maldict}. We have since added support for identifying malware family aliases using all three of our alias resolution approaches.

\subsection{Identifying Trivial Aliases}
\label{sec:trivial-alias}

If a token can be transformed into a second token by appending a single character (e.g. "Backdoor" to "Backdoor0"), and both tokens belong to the same lexical category, ClarAVy considers the pair to be trivial aliases. Additionally, ClarAVy uses a small list of common substrings to identify trivial aliases. If two tokens are identical except for one of those substrings at the beginning or end of a token, they become trivial aliases. For example, the families Kronos and Kronosbot would be considered trivial aliases due to the "-bot" substring that often appears at the end of family tokens.

\subsection{Identifying Sibling Aliases}
Malware family aliases can have very different spellings, making identification difficult. For example, the Andromeda, Gamarue, and Wauchos families are aliases. We say that two family tokens are sibling aliases if they both frequently co-occur with each other in AV scan data. ClarAVy does not search for sibling aliases between non-family tokens. To identify sibling aliases, we adapt AVClass' alias resolution methodology \cite{avclass}. 
Let the number of scan reports containing token $t_i$ be given by $|t_{i}|$, and let $|(t_{i}, t_{j})|$ be the number of scan reports containing both tokens $t_i$ and $t_j$. Then, the co-occurrence percentage of $t_i$ with $t_j$ is given by:

\begin{equation*}
\coocur(t_{i}, t_{j}) = \frac{|(t_{i}, t_{j})|}{|t_{i}|}
\end{equation*}\

ClarAVy accepts parameters $S$ (default $S$=0.95) and $T$ (default $T$=1000) to control sibling aliasing. If $\min(|t_i|, |t_j|) > T$ and \linebreak $\min(\coocur(t_{i}, t_{j}), \coocur(t_{j}, t_{i})) > S$, then $t_{j}$ and $t_{i}$ become sibling aliases. 

ClarAVy's methodology enforces a "two-way" relationship where both tokens must co-occur with each other at a high rate to become sibling aliases. This is a slightly different approach from AVClass, which takes the maximum of the co-occurrence percentages rather than the minimum, permitting pairs of tokens with just a "one-way" co-occurrence relationship to become aliases \cite{avclass}. However, when the difference between $\coocur(t_{i}, t_{j})$ and $\coocur(t_{j}, t_{i})$ is large, aliasing errors are more likely to occur.

\subsection{Identifying Parent-Child Aliases}
\label{sec:parent-child}
ClarAVy uses a third class of aliases, called parent-child aliases, to address pairs of tokens that have a "one-way" co-occurrence relationship. To become parent-child aliases, the less common token (the child token) must co-occur with the more common token (the parent token) in a sufficient percentage of scan reports. To reduce the number of aliasing errors, the two tokens must also be spelled similarly. We quantify this using a similarity score based on edit distance, defined as: 

\vspace*{-8pt}

\begin{equation*}
    \escore(t_{i}, t_{j}) = 1 - \edist(t_{i}, t_{j}) \; /  \; \min(\len(t_{i}), \len(t_{j}))
\end{equation*}\

ClarAVy accepts parameters $E$ (0.6 by default) and $C$ (0.5 by default) to control parent-child aliasing. Let $t_i$ be the child token and $t_j$ be the parent token. If $\escore(t_{i}, t_{j}) >= E$ and $\coocur(t_{i}, t_{j}) \times \escore(t_{i}, t_{j}) >= C$, then $t_{i}$ and $t_{j}$ become parent-child alias candidates.

\section{ClarAVy Tagging and Family Labeling}
\label{sec:family-labeling}

We will now detail several design choices and algorithmic improvements to support family labeling with greater accuracy than prior tools, with the ability to scale to enormous malware corpora sizes. After generating the token taxonomy and resolving aliases, ClarAVy reprocesses each AV scan report. This time, tokens that were previously assigned PRE or UNK may receive a more informative lexical category. Additionally, tokens with known aliases are replaced with their canonical names. ClarAVy aggregates the tokens from each AV scan report into a malware family label and a list of descriptive tags. Section \ref{sec:non-family-ranking} will discuss how ClarAVy tags malware, and Section \ref{sec:family-labeling-sub} will discuss ClarAVy's Variational Bayesian approach to family labeling. Before this, we must address how relationships between AV products affect tagging and family labeling.

\subsection{Relationships Between AV Products}
\label{sec:av-correlations}

\begin{figure}[!t]
    \centering
\includegraphics[width=.9\columnwidth,keepaspectratio]{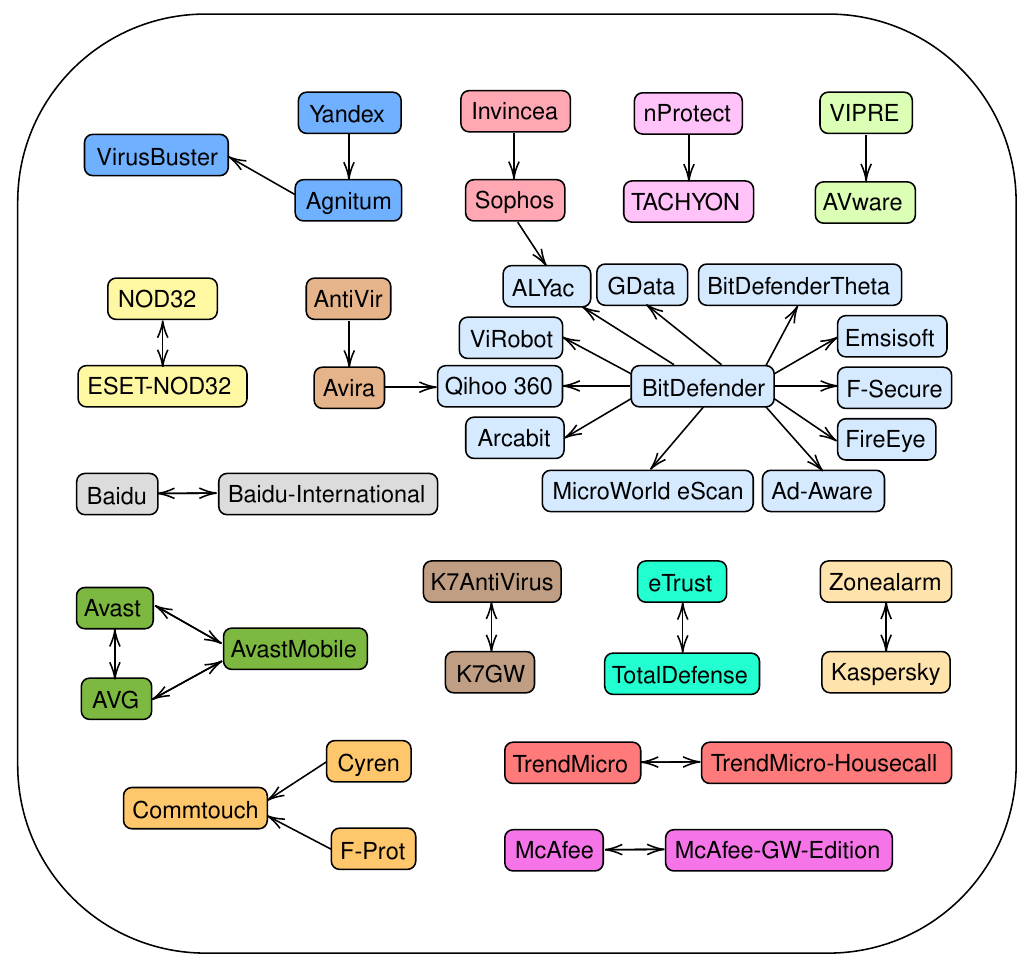}
    \vspace*{-6pt}
    \caption{Known relationships between AV products \cite{joyce2023maldict}\vspace*{-10pt}}
    \label{fig:av-correlation}
\end{figure}

The existence of relationships between AV products is well known in the malware analysis industry \cite{zhu2020, botacin2022antiviruses}. The leading causes include AV products sub-licensing their engines to others, AV products owned by the same company, and AV products copying other products' detection results \cite{avclass, av-meter}.  AV products with these relationships can produce correlated detections and these correlations must be taken into account during tagging and family labeling. There seem to be other factors contributing to these correlations as well, but they are poorly understood \cite{joyce2021rank1}. The clusters in Figure \ref{fig:av-correlation} show groups of AV products supported by ClarAVy that are known to be related due to ownership, acquisition, engine licensing, or sharing agreement \cite{joyce2023maldict}. ClarAVy attempts to account for these correlations while aggregating AV detections into tags and family labels. If two or more correlated AV products output the same token, they are treated as a single "vote" for their respective tag or family.

\subsection{Tagging Methodology}
\label{sec:non-family-ranking}
 For each AV scan report, ClarAVy counts the number of times each BEH, FILE, PACK, VULN, and GRP token occurred in the report. If a token occurs at least $M$ times in the outputs of unrelated AV products (where $M$ is a threshold parameter), then that token will become a tag associated with that malicious file. $M$ can be set for each individual token type. By default $M=5$ for BEH and FILE tokens and $M=1$ for VULN, PACK, and GRP tokens. These defaults were empirically selected in our prior work \cite{joyce2023maldict}. For each tag, ClarAVy outputs the number of unrelated AV products that contributed to that tag. When a token becomes a tag, it is normalized by converting it to lowercase and replacing all special characters with underscores.

\subsection{Family Labeling}
\label{sec:family-labeling-sub}
Unlike the tasks in Section \ref{sec:non-family-ranking} where multiple tags can be assigned to a malicious file, each malicious file can belong to (at most) one malware family. This allows us to adapt prior work on aggregating crowdsourced labels to our domain, resulting in higher accuracy than simple voting schemes. We base our approach on the Dawid-Skene algorithm, which has historically been used for aggregating the annotations of crowdsourced voters \cite{dawid-skene}. The Dawid-Skene algorithm assumes that $N$ data points have been annotated by $K$ annotators, with each annotation being one of $L$ possible classes. Annotators are permitted to abstain from voting. Dawid-Skene learns a confusion matrix of dimension $L$ x $L$ for each of the $K$ annotators, encoding (for each possible combination of classes) the likelihood that the annotator has mistakenly voted for one class rather than the correct one. For each of the $N$ data points, Dawid-Skene outputs a probability distribution over the $L$ classes.

The quadratic dependence of $L$ in the Dawid-Skene algorithm is not usually an issue because crowdsourcing tasks are not often in an extreme multiclass domain. However, the largest public malware datasets may contain tens of millions of files and tens of thousands of malware families \cite{sorel, virusshare}. 
In Section \ref{sec:large-scale}, we use ClarAVy to label 39,747,485 files from 52,371 malware families. 
Assuming $L$=50,000 families, $K$=103 AV product "annotators", and 32-bit floating point precision, the confusion matrices learned by the Dawid-Skene algorithm would consume more than 1TB of memory. Calculating posterior probabilities for $L$=50,000 families for each of $N$=40,000,000 files would require more than 8TB of memory. For Dawid-Skene to realistically be applied to a dataset of our required size, modifications to the algorithm are necessary. 

The Dawid-Skene algorithm has been extended in numerous other works \cite{carpenter2008multilevel,zhang2014spectral,sinha2018fast}. We adapt one such work by \citet{IBCC}, who introduces the Independent Bayesian Classifier Combination (VB--IBCC), a Variational Bayesian approach for quickly approximating the Dawid-Skene algorithm. We also closely follow the work of \citet{maximilien2017}, who created a sparse implementation of VB-IBCC that assumes $N$ (the size of the dataset) and $K$ (the number of annotators) are both extremely large, but $L$ (the number of classes) is relatively small. Unlike \citet{maximilien2017}, our use case has extremely large $N$ and $L$ and relatively small $K$. To support these conditions, we created our own implementation of VB-IBCC that supports sparse confusion matrices and sparse label probabilities. ClarAVy uses our sparse VB-IBCC implementation, which we call SparseIBCC, to efficiently and accurately label massive malware datasets.

\begin{figure}[h]
    \centering
\includegraphics[width=0.9\columnwidth,keepaspectratio]{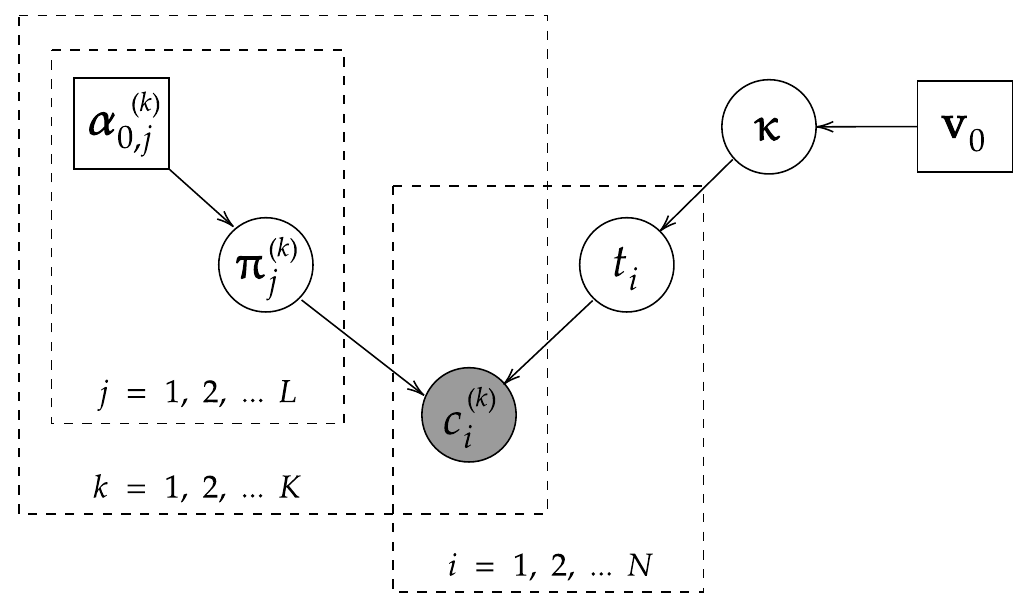}
    \caption[Plate diagram for VB-IBCC]{Plate Diagram for VB-IBCC \cite{IBCC}}
    \label{fig:ibcc_plate_diagram}
\end{figure}

We begin by defining the VB-IBCC statistical model \cite{IBCC}. The objective of VB-IBCC is to infer the latent class $t_i$ of each data point ($i = 1, 2, ... N$) and where $t_i$ is one of the valid class labels $j$ ($j = 1, 2, ... L$). VB-IBCC assumes that $t_i$ is from a Multinomial distribution where class probabilities are given by $\boldsymbol{\kappa} \; : \; p(t_i = j | \boldsymbol{\kappa}) = k_j$ \cite{IBCC}. The prior on $\boldsymbol{\kappa}$ is given by $\boldsymbol{\nu} = \left[\nu_{0,1} \; ... \; \nu_{0,L}\right]$.

The output of a single classifier $k$ is denoted $c_i^{(k)}$. VB-IBCC assumes that $c_i^{(k)}$ is generated from a Multinomial distribution that is dependent on $t_i$, denoted $\boldsymbol{\pi}_j^{(k)} \; : \; p(c_i^{(k)} = l| t_i = j, \boldsymbol{\pi}_j^{(k)}) =  \boldsymbol{\pi}_{jl}^{(k)}$ \cite{IBCC}. The hyperparameter for $c_i^{(k)}$ is $\boldsymbol{\alpha}_{0,j}^{(k)} \; = \; \left[\boldsymbol{\alpha}_{0,j1}^{(k)} \; ... \; \boldsymbol{\alpha}_{0,jL}^{(k)}\right]$. The set of Dirichelet distributions $\boldsymbol{\pi}_j^{(k)}$ for each annotator and for each class are denoted $\boldsymbol{\Pi} = \left\{ \boldsymbol{\pi}_j^{(k)} | j = 1 ... L, k = 1 ... K \right\}$, and the set of hyperparameters for these distributions are denoted $\boldsymbol{A_0} = \left\{\boldsymbol{\alpha}_{0,j}^{(k)} | j = 1 ... L, k = 1 ... K \right\}$ \cite{IBCC}. VB-IBCC learns to approximate the latent variables $\textbf{\textit{t}}$, $\boldsymbol{\Pi}$, and $\boldsymbol{\kappa}$ using Variational Bayes \cite{IBCC}. This involves an iterative procedure where each iteration approximates the E-step and the M-step of the Expectation Maximization algorithm, until convergence \cite{EM}. 

\subsection{Sparse Confusion Matrices in SparseIBCC}
Our SparseIBCC implementation enables the VB-IBCC algorithm to support an extreme number of classes. Recall that $\boldsymbol{\pi}^{(k)}$ represents the confusion matrix of $k^{th}$ annotator in $\boldsymbol{\Pi}$. If families $j$ and $l$ never co-occur in any AV scan reports, then there is no evidence that family $j$ has been confused for family $l$ or vice versa. Because of this, $\boldsymbol{\pi}_{jl}^{(k)}$ and $\boldsymbol{\pi}_{lj}^{(k)} = 0, \; \forall k=1,2,...K$.

We observe that most malware families only co-occur with a very small subset of other families. For example, we found that $\approx$99.5\% of possible malware family pairs never co-occur within AV scan reports in the MOTIF dataset \cite{MOTIF}. This property permits $\boldsymbol{\Pi}$, as well as its prior $\boldsymbol{A_0}$, to be represented using a custom sparse format. Let $F_j$ be an ordered set of malware families that co-occur with family $j$. Then, each $\boldsymbol{\alpha}_0,j^{(k)} \in \boldsymbol{A_0}$ and each $\boldsymbol{\pi}_j^{(k)} \in \boldsymbol{\Pi}$ can be represented as arrays of length $|F_j|$, where $|F_j| \leq L$ (and typically $|F_j| << L$). To avoid the use of ragged arrays, SparseIBCC stores $\boldsymbol{\Pi}$ and $\boldsymbol{A_0}$ as matrices of shape $\sum_{j=1}^L |F_j| \; \times \; K$, and an additional data structure is used to navigate these matrices. Under the assumption that $\approx$99.5\% of possible malware family pairs do not co-occur in AV scan reports, SparseIBCC uses $\approx$200$\times$ less memory to store $\boldsymbol{\Pi}$ and $\boldsymbol{A_0}$ than a dense implementation.

\subsection{Sparse Probabilistic Labels in SparseIBCC}

Computing and storing $t$ (the posterior) for $N$ data points over all $L$ classes is intractable when $N$ and $L$ have extreme size. Therefore, we assume that $t_i \in c_i$; that is, one of the annotated labels is the correct label. Although this is a na\"ive assumption in the general case, it is a reasonable one in the malware labeling domain. We expect malicious files to be scanned by a set of $\approx$70 AV products from VirusTotal. If none of those AV products have a particular family in their detections, then we have extremely little confidence that that family is the correct one.

The $O(NL)$ memory consumption of VB-IBCC for storing the posterior becomes $O(NK)$ with this alteration. In our scenario where $L$=50,000 and $K$=103, our sparse approach would still consume $\approx$500$\times$ less memory than VB-IBCC for storing $t$ in the worst case. If we conservatively assume that each AV scan report contains 10 distinct families on average, this becomes $\approx$5,000$\times$ less memory consumption.  In experiments with small-scale malware datasets we found this change also increased accuracy.

\subsection{Family Confidence Scores}

We found that after convergence, the posterior probability estimate of the most likely family for each scan was often very close to one, while all other posterior probabilities tended to be near zero. We believe that this is due to an interaction between the Softmax function and the extremely multiclass nature of our data. Because of this, for each scan report, ClarAVy outputs the most likely family and a confidence score determined by a separate model. Each confidence score indicates the "difficulty" of accurately labeling a given scan report. In general, we observe that scan reports with few AV detections, few malware families, and/or high disagreement between AV products are the most difficult to correctly label. The following features are used for difficulty estimation:

\begin{itemize}
    \item The number of distinct families that were detected.
    \item The Shannon's entropy the detected families.
    \item The number of malicious detections divided by the total number of AVs that scanned the file.
    \item The number of AV labels with family names divided by the total number of detections.
    \item The number of AV labels with family names divided by the number of AVs that scanned the file.
    \item The probability of SparseIBCC's most likely family.
    \item The Shannon's entropy of SparseIBCC's predictions.
\end{itemize}

We trained an XGBoost classifier on a combination of the MOTIF and MalPedia dataset (with duplicates removed) using five-fold cross-validation. During training, the classifier was given the features listed above as input, and it predicted the probability that SparseIBCC made the correct family prediction. ClarAVy's confidence scores allow practitioners to use only high-confidence family labels, if that is a requirement. We provide a study of ClarAVy's family labeling accuracy relative to confidence score in Section \ref{sec:eval-threshold}.

\section{Experiments}
\label{sec:evaluation}

In our prior work, we evaluated ClarAVy's AV parsing, alias resolution, and tagging performance \cite{joyce2023maldict}. The focus of our evaluation in this work is on the malware family labeling task. 
During our evaluation, we compare ClarAVy with the five tools in Section \ref{sec:related_work}. We also study how ClarAVy's accuracy is affected by confidence score. Finally, we show that ClarAVy can scale to a dataset of $\approx$40 million VirusTotal scan reports, and we investigate the token taxonomy and alias mapping that ClarAVy generated from that dataset. 

\subsection{Family Labeling Experiments}
\label{sec:comparing-claravy}

Our evaluation is performed using the MOTIF and MalPedia datasets. MOTIF contains 3,095 malicious PE files with ground truth family labels \cite{MOTIF}. MalPedia is continuously being updated and includes malware from a variety of file formats \cite{MalPedia}. Some files in MalPedia have not been uploaded to VirusTotal and were excluded from our evaluation. We also performed label cleaning on MalPedia to remove generic family names and to fix unresolved family aliases. After this process, we were left with 9,847 files from 2,755 families in MalPedia. Importantly, we selected the MOTIF and MalPedia datasets because they have high-quality labels that are not derived from any antivirus products. MOTIF is labeled using open source reporting, and MalPedia is labeled using a mix of open-source reporting and YARA rule detections \cite{MOTIF, MalPedia}.

Each tool in our evaluation has its own preferences for family alias naming. For the purposes of computing accuracy, a tool is considered to have made a correct prediction if it outputs any alias of the true malware family. Family prediction accuracies are shown in Table \ref{tab:motif-family-eval}, where ClarAVy dominates all prior methods, and an ablation shows our SparseIBCC improves accuracy by $\approx$4-5\% over plurality voting. For each tool, we describe our experimental procedure and discuss the results.

\vspace*{-4pt}

\begin{table}[H]
\centering
\caption{Comparing ClarAVy's family labeling accuracy against prior tools. ClarAVy is 8 percentage points more accurate than all other tools on MOTIF, and 12 percentage points more accurate than all other tools on MalPedia.}
\adjustbox{max width=\columnwidth}{

\begin{tabular}{@{}lrr@{}}

\toprule
Labeler Name & MOTIF  & MalPedia  \\
\midrule
AVClass (Defaults) & 51.41\% & 43.78\%\\ 
AVClass w/ Update Module & 54.09\% & 44.08\%\\
EUPHONY & 29.23\% & OOM\\  
AVClass++ (Defaults) & 44.09\% & 41.07\% \\
Sumav & DNF & DNF \\
TagClass w/ Plurality Voting & 37.38\% & 33.89\% \\
TagClass w/ Dawid-Skene & $\approx$56\% \textsuperscript{1} & N/A\\
ClarAVy w/ Plurality Voting & 60.00\% & 51.23\% \\
\textbf{ClarAVy w/ SparseIBCC} & \textbf{64.16\%} & \textbf{56.88\%} \\

\bottomrule
\end{tabular}
}
\label{tab:motif-family-eval}
\end{table}

\footnotetext[1]{Unable to reproduce family inference accuracy experiments from \cite{tagclass_DS} because code is not available.}

\subsubsection{Evaluating AVClass} AVClass comes with a default token taxonomy and alias mapping. It also has an update module that allows a token taxonomy and alias mapping to be generated from a large corpus of VirusTotal scans \cite{avclass}. We ran the update module using our collection of $\approx$40 million VirusTotal scans. Compared to the default settings, this increased AVClass' accuracy by 2.68 percentage points on MOTIF and 0.30 percentage points on MalPedia. 

We found that AVClass' update module has a high false positive rate when identifying aliases, primarily due to downstream errors during AV parsing. When run on our set of $\approx$40 million scan reports, AVClass' update module found 4,117 new pairs of token aliases. However, 1,902 of these alias pairs included at least one token which ClarAVy identifies as a suffix token. At minimum, this is an error rate of 46.20\%. 
In a manual inspection of the new alias pairs, we found additional errors that were likely caused by AVClass' loose co-occurrence requirements during alias resolution.

\subsubsection{Evaluating EUPHONY}
    We ran EUPHONY under default settings on both datasets \cite{euphony}. EUPHONY's accuracy on the MOTIF dataset was just 29.23\%. When run on the MalPedia dataset, EUPHONY consumed 128GB of RAM and crashed with an "Out of Memory" (OOM) error. Running out of memory on fewer than 10,000 scans demonstrates EUPHONY is unable to scale to large dataset sizes, and other settings do not alleviate this constraint.

\subsubsection{Evaluating AVClass++}
    AVClass++ is a fork of the original version of AVClass, prior to AVClass' merge with AVClass2 \cite{avclassplusplus, avclass, avclass2}. When run with default settings, AVClass++ is outperformed by the current version of AVClass, likely due to improvements made when the original AVClass was combined with AVClass2.

\subsubsection{Evaluating Sumav}
    Sumav creates a token taxonomy by building a graph of related tokens in scan reports \cite{sumav}. We attempted to run Sumav's graph building module on both datasets, but after 24 hours neither program had completed. These runtime requirements make Sumav impractical to use in real-world settings.

\subsubsection{Evaluating TagClass}
    We ran TagClass' update, clean, and parse modules on MOTIF and MalPedia with default settings to generate a token taxonomy. Potential family tokens given a score greater than or equal to 8 (out of a possible 10) were accepted as family names. 
    We then used plurality voting to infer malware families for the files in MOTIF and MalPedia. TagClass with plurality 
    voting was 37.38\% accurate on MOTIF and 33.89\% accurate on MalPedia. The primary sources of error in TagClass' predictions were suffix tokens being incorrectly identified as family tokens and a lack of alias resolution.

    Unfortunately, the code implementing TagClass' Dawid-Skene malware family inference is not publicly available, so we were unable to reproduce experiments by \citet{tagclass_DS}. The authors reported an accuracy of $\approx$56\% on 1,972 files selected from MOTIF, which is $\approx$19 percentage points higher than our plurality voting findings for TagClass. The difference in voting strategy is unlikely to be the sole cause of this discrepancy. When evaluating TagClass on MOTIF, \citet{tagclass_DS} excluded files for which TagClass identified no families. Additionally, they excluded entire malware families from MOTIF at their discretion. We believe that these actions may have inflated their reported accuracy due to selection bias.

\subsubsection{Evaluating ClarAVy}
\label{sec:claravy_motif_MalPedia}

We first labeled MOTIF and Malpedia with ClarAVy using plurality voting rather than SparseIBCC. This allowed us to compare ClarAVy against other AV labeling tools that use plurality voting, such as AVClass and AVClass++. On average, ClarAVy was 6.53 percentage points more accurate than AVClass (with update module), 13.04 percentage points more accurate than AVClass++, and 19.98 percentage points more accurate than TagClass when each tool used plurality voting. This suggests that ClarAVy has higher-fidelity AV parsing and alias resolution than the other surveyed tools, since the voting method was constant.

We then labeled both datasets using ClarAVy with default settings. The use of SparseIBCC rather than plurality voting raised ClarAVy's accuracy by an average of 4.91 percentage points. ClarAVy was 11.44 percentage points more accurate than AVClass (with update module) on average. It was approximately 8 percentage points higher than TagClass' reported accuracy on MOTIF.

\subsection{Family Confidence Score Experiments}
\label{sec:eval-threshold}

\begin{figure}[h!]
    \centering
\includegraphics[width=0.925\columnwidth,keepaspectratio]{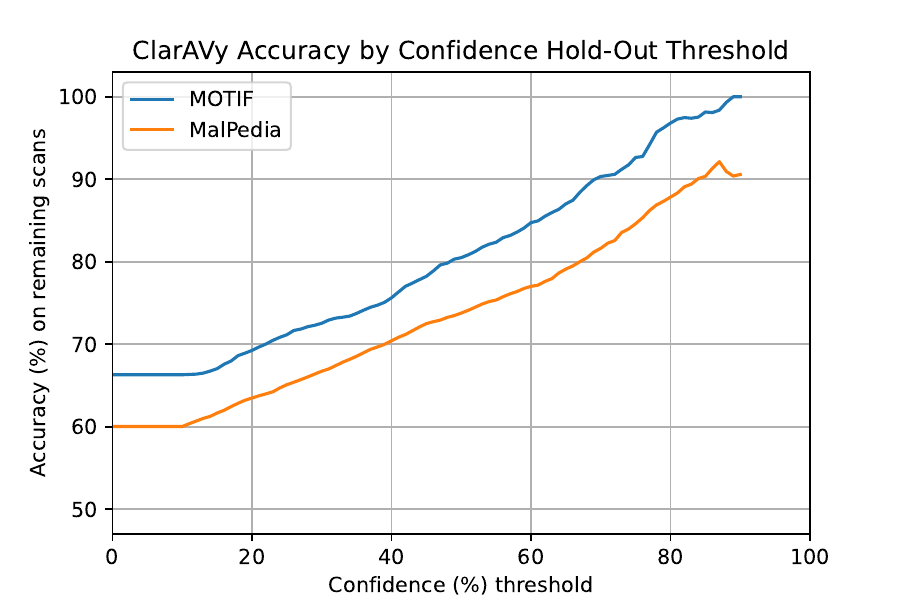}
    \caption[]{Accuracy of ClarAVy if scans below the given confidence are ignored. No scans had a confidence of 92\% or higher. ClarAVy has more than 90\% accuracy on MOTIF when using a confidence threshold of 70\%.}
    \label{fig:confidence-threshold}
    \vspace*{-8pt}
\end{figure}

\begin{figure}[h!]
    \centering
\includegraphics[width=0.925\columnwidth,keepaspectratio]{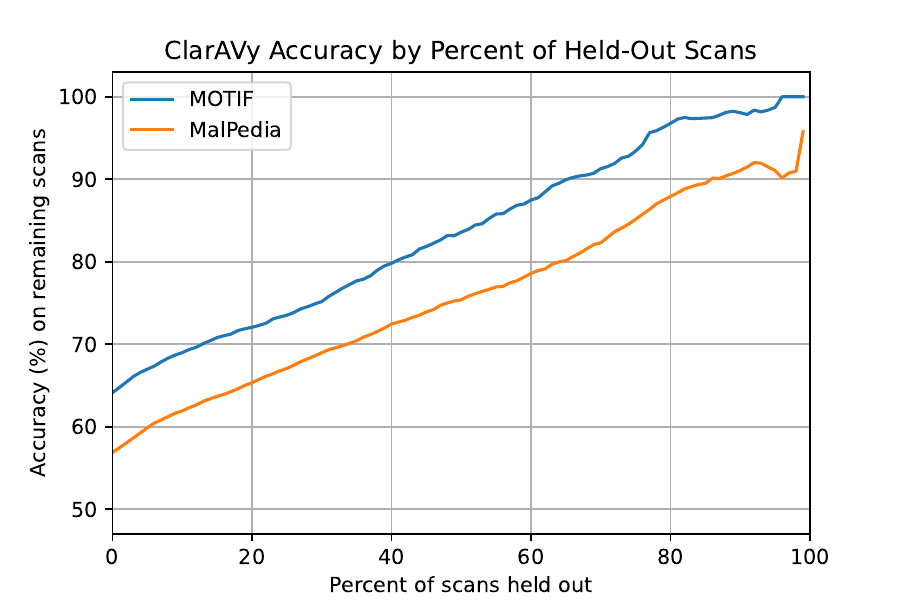}
    \caption[]{Accuracy of ClarAVy if a given percentage of scans with the lowest confidence scores are held out. ClarAVy's accuracy increases almost linearly as scans with the next-lowest confidence are held out.}
    \label{fig:confidence-pct}
\end{figure}

ClarAVy provides a confidence score for each of its predictions. Users who want only high-confidence family labels can ignore family labels that recieve a confidence score below a given threshold. For example, Figure \ref{fig:confidence-threshold} shows that ClarAVy labels for the MOTIF dataset with a confidence score of 70\% or above are $\approx$90\% accurate. Figure \ref{fig:confidence-pct} shows that our confidence scores are stable -- as family labels with the next-lowest confidence scores are ignored, the accuracy of ClarAVy's family inference on the remaining scan reports increases close to linearly. 

Practitioners that build malware datasets by discarding files with low-confidence scan reports are cautioned to consider selection bias in their methodology. Family labels with higher-confidence scores are likely to correspond to malicious files that are less evasive and/or have better AV coverage \cite{li}. 

\subsection{Running ClarAVy on Large-Scale Datasets}
\label{sec:large-scale}

To show that ClarAVy can scale to massive dataset sizes, we ran it on the $\approx$40 million VirusTotal scans we collected for VirusShare chunks 0-465, passing flags to output the generated taxonomy and alias mapping \cite{virusshare, virustotal}. While processing this dataset, ClarAVy identified 48,417 malware families which were not part of its default token taxonomy and resolved 4,472 pairs of aliases which were not in its default alias mapping. More statistics about ClarAVy's lexical category assignments are shown in Table \ref{tab:taxonomy-stats}. A list of the most common families and tags output by ClarAVy is in Appendix A.

\begin{table}[h!]
\caption{The number of tokens of each type in ClarAVy's default taxonomy, and the number of tokens that were identified by running ClarAVy on $\approx$40 million VirusTotal reports.}
\label{tab:taxonomy-stats}
\begin{tabular}{@{}lrrr@{\hskip 12pt}l@{}}
\toprule
Type & Default Tax. & Generated Tax. & New Tokens\\
\midrule
FAM & 3,973 & 52,371 & 48,398 \\
GRP & 240 & 258 & 18 \\
BEH & 2,626 & 3,092 & 466 \\
FILE & 381 & 472 & 91 \\
PACK & 144 & 145 & 1 \\
HEUR & 93 & 2,798 & 2705 \\
PRE & 193 & 1,619 & 1,426 \\
\bottomrule
\end{tabular}
\end{table}

In the former experiment, ClarAVy ran on Intel Xeon E7-8870 CPUs with 128 threads. ClarAVy spent 17 hours and 44 minutes parsing the set of scan reports for the first time, 35 minutes generating the token taxonomy and performing alias resolution, 18 hours and 49 minutes on the second pass over the scan reports, and 1 hour and 3 minutes on SparseIBCC. Disk I/O is ClarAVy's main bottleneck, and we believe that reading scan reports over a network-mounted drive caused AV parsing to be especially slow in this instance. SparseIBCC took up just a portion of ClarAVy's overall runtime because it has been optimized with just-in-time compilation and benefits from the high number of threads used.

\begin{figure}[!h]
\vspace*{-8pt}
    \centering
\includegraphics[width=\columnwidth,keepaspectratio]{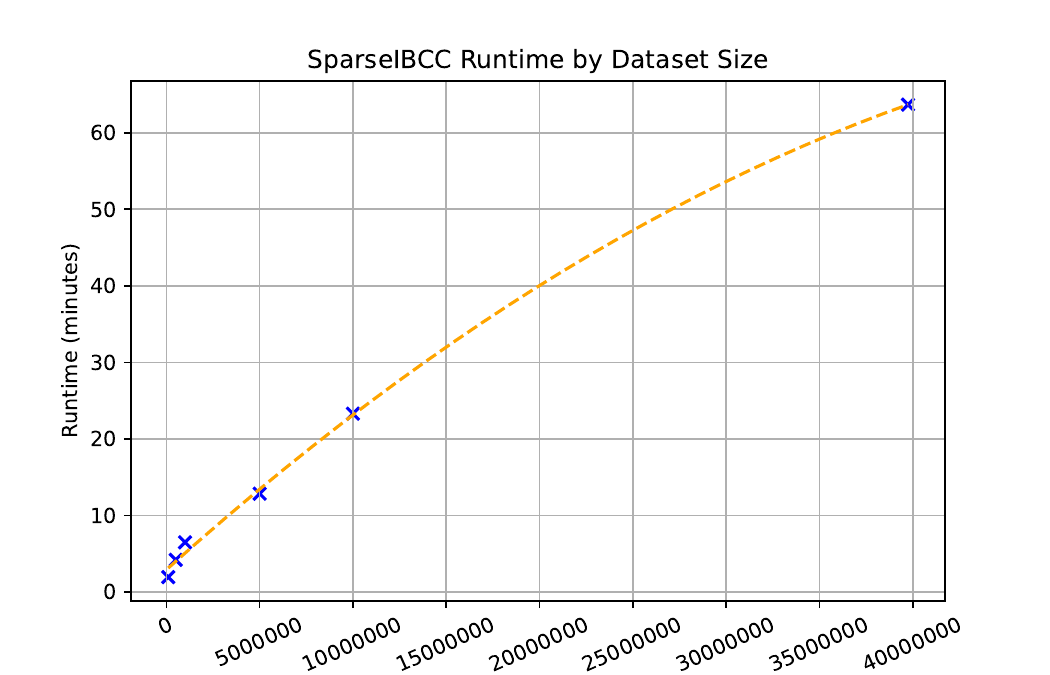}
    \caption[]{SparseIBCC runtime in comparison to dataset size.}
    \label{fig:runtime}
\end{figure}

To investigate how SparseIBCC runtime scales with the number of AV scan reports, we re-ran ClarAVy on the first 100K, 500K, 1M, 5M, and 10M VirusTotal reports from our VirusShare collection. We plotted the runtime of SparseIBCC in Figure \ref{fig:runtime}. The figure also includes a quadratic curve of best fit for the collected data, which indicates that SparseIBCC's runtime complexity is sub-linear in practice. We expect that this is because the rate of encountering new families decreases as dataset size increases. ClarAVy is able to scale to much larger dataset sizes than TagClass (which is bound by the quadratic memory requirements of the original Dawid-Skene algorithm), EUPHONY (which ran out of memory on $\approx$10,000 files), and Sumav (which took more than 24 hours to process $\approx$3,000 files).

\vspace*{-4pt}

\subsection{Family Labeling Discussion}
\label{sec:discussion}

Family classification is an inherently difficult task due to the unreliability of AV detections, naming inconsistencies, and its extremely multiclass nature. Although ClarAVy is an improvement over prior work, its base accuracy may still not be sufficient for applications that require extremely precise family labels. There is diminishing opportunity to further raise base accuracy due to the overall difficulty of scan reports that ClarAVy mislabels. In MOTIF, 18.64\% of scan reports are impossible to correctly label using AV aggregation because no AV detections contain the true family name. An additional 11.53\% of scan reports are extremely difficult to label because the true family is only present in a single AV detection.

Encouragingly, many of the incorrectly predicted labels in our evaluation of ClarAVy were somehow related to the true family. For example, ClarAVy incorrectly labeled 168 files in MalPedia as the Zeus family. However, 133 of these (79\%) were from the VMZeus, Citadel, KINS, Zeus Panda, Zeus Sphinx, Floki Bot, or Gameover Zeus families. Each of these are variants of Zeus; they are based on Zeus' source code but are different enough to receive a new separate family name. In another instance, ClarAVy incorrectly labeled 47 files in MalPedia as the NukeSped family. NukeSped is a remote access tool that is a signature family of the Lazarus group, which is a North Korean APT group. We found that the 47 files incorrectly labeled as NukeSped belonged to 33 other families, all of which were attributed to a North Korean threat group (one family to Kimsuky, and the rest to the Lazarus group). AV products seem to be using NukeSped as a catchall when they cannot determine a more specific family for malware associated with North Korea. 

In these case studies, the lack of coverage and/or specificity by AV products led to ClarAVy outputting an incorrect family name. We observe that AV products prioritize malicious detection while accurate family identification is a secondary concern. To make significant progress possible, AV products must devote more resources to improved family coverage.

\section{Industry Use and Conclusion}
\label{sec:conclusion}

ClarAVy has already been in an industry setting to label more than 1.5 million VirusTotal scans collected between November 2023 and April 2024. We have confirmed that although there are minor drifts in AV detection structure and naming within this newer set of data, ClarAVy's AV parsing functions remain effective and robust. We are actively working on deploying ClarAVy to a live feed of malware, where it will label and tag $\approx$400,000 files daily.

The prior version of ClarAVy is available on GitHub\textsuperscript{2} as of the time of writing \cite{joyce2023maldict}. Following publication, ClarAVy will be updated to include the contributions of this work: support for accurate and scalable malware family labeling, as well as incremental improvements to AV parsing, alias resolution, and tagging. We plan to continue supporting ClarAVy and are actively working on new features, such as improved confidence scoring.

\footnotetext[2]{https://github.com/FutureComputing4AI/ClarAVy/}

To our knowledge, ClarAVy is the first AV-based malware labeling tool with an intelligent aggregation strategy that can scale to datasets in the tens of millions of scan reports. We found that ClarAVy's malware family labels were an average of 11.44 percentage points more accurate than AVClass'. ClarAVy will facilitate improvements in downstream tasks that require large quantities of high-quality malware family labels, such as the training of ML classifiers.

\bibliographystyle{ACM-Reference-Format}
\balance
\bibliography{sample-base}

\onecolumn
\newpage
\begin{appendices}
\section{}

\begin{table}[!h]
\caption{The 50 most common families and tags applied to VirusShare chunks 0-465, descending.}
\adjustbox{max width=\columnwidth}{
\begin{tabular}{@{}llllll@{\hskip 12pt}l@{}}
\toprule
FAM & GRP & BEH & FILE & VULN & PACK\\
\midrule
ramnit & equationgroup & downloader & js & ms04\_028 & nsis\\
blackhole & turla & iframe & pdf & cve\_2017\_11882 & upx\\
coinhive & bronzebutler & backdoor & msil & cve\_2017\_0199 & themida\\
faceliker & darkhotel & adware & vbs & ms03\_43 & vmprotect\\
fbjack & strongpity & worm & html & cve\_2012\_1723 & aspack\\
virut & apt28 & redirector & android & cve\_2007\_0071 & upack\\
multiplug & apt1 & dropper & dos & ms06\_014 & fsg\\
installcore & apt32 & virus & bat & cve\_2006\_4777 & nsanti\\
softpulse & lazarusgroup & passwordstealer & autoit & cve\_2014\_6332 & pecompact\\
firseria & apt29 & spyware & w97m & cve\_2010\_1885 & nspack\\
cryxos & gamaredon & startpage & java & cve\_2018\_0870 & nspm\\
loadmoney & apt12 & fakeantivirus & php & cve\_2008\_2551 & mystic\\
zeus & lamberts & injector & lnk & cve\_2017\_1182 & mpress\\
onlinegames & ta505 & exploit & win64 & cve\_2011\_2462 & rlpack\\
domaiq & patchwork & dialer & linux & cve\_2010\_0806 & mew\\
wacatac & muddywater & pua & script & cve\_2012\_0158 & molebox\\
vobfus & apt27 & packed & powershell & cve\_2010\_2568 & execryptor\\
installrex & operationtransparenttribe & rootkit & win95 & ms05\_009 & armadillo\\
allaple & hellsing & banker & macosx & ms08\_067 & pespin\\
kykymber & operationirontiger & phishing & macroword & cve\_2014\_6352 & ntkrnlpacker\\
codecpack & apt17 & coinminer & hllo & cve\_2014\_0496 & privateexeprotector\\
hupigon & operationnightdragon & autorun & shellcode & cve\_2014\_0545 & lighty\\
morstar & apt33 & ransom & x97m & ms06\_024 & zprotect\\
megasearch & apt36 & clicker & python & cve\_2012\_0507 & orien\\
solimba & operationlotusblossom & bho & autocad & cve\_2010\_4452 & pearmor\\
sality & fin7 & antiav & perl & cve\_2013\_1331 & asprotect\\
twitscroll & greenbug & ircbot & win16 & cve\_2009\_2501 & yoda\\
c99shell & unit8200 & hacktool & swf & ms04\_025 & exestealth\\
yuner & operationdarkseoul & proxy & excelformula & cve\_2012\_1889 & pex\\
redirme & operationgrizzlysteppe & keylogger & j2me & cve\_2010\_0188 & expressor\\
bifrose & apt10 & installer & symbos & cve\_2012\_1526 & bero\\
qqrob & apt35 & flooder & macroexcel & cve\_2021\_26855 & enigmaprotector\\
electrorat & sandworm & riskware & unix & cve\_2019\_0903 & npack\\
overdoom & platinum & fakealert & vba & cve\_2014\_6342 & packman\\
renos & tontoteam & fraud & msword & cve\_2010\_3962 & obsidium\\
scriptip & donotteam & remoteshell & msexcel & cve\_2017\_8759 & thinstall2425\\
blinkx & ke3chang & patcher & driver & cve\_2017\_2927 & telock\\
outbrowse & operationdesertfalcons & fakejquery & rtf & cve\_2010\_0840 & vprotect\\
emotet & apt34 & cryptor & msoffice & ms03\_043 & maskpe\\
buzus & apt40 & joke & asp & cve\_2011\_3544 & xcomp\\
fbscam & apt37 & bootkit & multi & cve\_2013\_2423 & bitarts\\
tdss & apt15 & blocker & macro & cve\_2012\_4681 & spack\\
optimuminstaller & easternroppels & keygen & autolisp & cve\_2021\_27065 & mpack\\
nemucod & molerats & toolbar & a97m & cve\_2017\_0245 & exe32pack\\
xmrig & operationtropictrooper & constructor & delphi & cve\_2011\_0611 & repacked\\
refroso & apt30 & hoax & comwar & cve\_2010\_2586 & niceprotect\\
flystudio & blacktech & regrun & ruby & cve\_2013\_0028 & upc\\
archsms & volatilecedar & servstart & rar & cve\_2010\_3333 & pcmm\\
bicololo & ta428 & killfiles & corrupted & cve\_2020\_0601 & nakedpack\\
fundf & ta21 & antifw & winnt & cve\_2014\_2804 & ppp\\

\bottomrule
\end{tabular}
}
\end{table}

\end{appendices}

\end{document}